\documentclass{article}

\usepackage{PRIMEarxiv}

\usepackage[utf8]{inputenc} 
\usepackage[T1]{fontenc}    
\usepackage{hyperref}       
\usepackage{url}            
\usepackage{color}
\usepackage{booktabs}       
\usepackage{amsfonts}       
\usepackage{nicefrac}       
\usepackage{microtype}      
\usepackage{lipsum}
\usepackage{fancyhdr}       
\usepackage{graphicx}       
\usepackage{amsmath}
\usepackage{makecell}
\usepackage[ruled,vlined,linesnumbered]{algorithm2e}
\usepackage{upgreek}
\usepackage{mhchem}
\usepackage[backend=biber,style=numeric,maxnames=3,minnames=1,sorting=none]{biblatex}
\newcommand{\micron}{$\upmu$m}
\newcommand{\rvec}{\boldsymbol{r}}
\newcommand{\fvec}{\boldsymbol{f}}
\newcommand{\xvec}{\boldsymbol{x}}
\newcommand{\cvec}{\boldsymbol{c}}
\newcommand{\norm}[1]{\left\lVert#1\right\rVert}
\newcommand{\admmpenalty}{\tau}
\newcommand{\degree}{^\circ}
\newcommand{\npr}{N_{\mathrm{inner}}}
\DeclareMathOperator*{\argmin}{arg\,min}

\title{Fidelity-preserving enhancement of ptychography with foundational text-to-image models}

\author{
  Ming Du$^*$, Volker Rose, Junjing Deng \\
  Advanced Photon Source \\
  Argonne National Laboratory \\
  Lemont, IL, USA \\
  $^*$ \texttt{mingdu@anl.gov} \\
  \AND
  Dileep Singh \\
  Applied Materials Division \\
  Argonne National Laboratory \\
  Lemont, IL, USA \\
  \And
  Si Chen, Mathew J. Cherukara$^\dagger$ \\
  Advanced Photon Source \\
  Argonne National Laboratory \\
  Lemont, IL, USA \\
  $^\dagger$ \texttt{mcherukara@anl.gov} \\
}

\begin{document}

\textbf{GOVERNMENT LICENSE}

The submitted manuscript has been created by UChicago Argonne, LLC, Operator of Argonne
National Laboratory (``Argonne''). Argonne, a U.S. Department of Energy Office of Science laboratory, is operated under Contract No. DE-AC02-06CH11357. The U.S. Government retains for
itself, and others acting on its behalf, a paid-up nonexclusive, irrevocable worldwide license in
said article to reproduce, prepare derivative works, distribute copies to the public, and perform
publicly and display publicly, by or on behalf of the Government. The Department of Energy will
provide public access to these results of federally sponsored research in accordance with the DOE
Public Access Plan. http://energy.gov/downloads/doe-public-access-plan.
\newpage

\maketitle

\begin{abstract}
Ptychographic phase retrieval enables high-resolution imaging of complex samples but often suffers from artifacts such as grid pathology and multislice crosstalk, which degrade reconstructed images. We propose a plug-and-play (PnP) framework that integrates physics model-based phase retrieval with text-guided image editing using foundational diffusion models. By employing the alternating direction method of multipliers (ADMM), our approach ensures consensus between data fidelity and artifact removal subproblems, maintaining physics consistency while enhancing image quality. Artifact removal is achieved using a text-guided diffusion image editing method (LEDITS++) with a pre-trained foundational diffusion model, allowing users to specify artifacts for removal in natural language. Demonstrations on simulated and experimental datasets show significant improvements in artifact suppression and structural fidelity, validated by metrics such as peak signal-to-noise ratio (PSNR) and diffraction pattern consistency. This work highlights the combination of text-guided generative models and model-based phase retrieval algorithms as a transferable and fidelity-preserving method for high-quality diffraction imaging.
\end{abstract}


\section{Introduction}

Coherent diffraction imaging (CDI) techniques have become indispensable for exploring the microscopic world since their inception. By illuminating a specimen with a coherent beam of light or electrons and directly collecting the far-field diffraction patterns of the beam passing through the specimen without an objective lens, one can computationally retrieve the phase modulation of the sample as the solution of an inverse problem, which process is known as phase retrieval \cite{Fienup1982-ap} or image reconstruction. Through the elimination of objective lenses, lens-induced aberrations are avoided in CDI, and the imaging resolution is no longer limited by the collection angle of the objective. Also, through the computational retrieval of the phase, low-absorption samples such as biological tissues can be imaged with much better contrast compared to an absorption-contrast imaging. The use of bright x-ray beams, often produced at synchrotron light sources, further allows one to image thick samples non-intrusively with high resolution \cite{Jacobsen2019-iq}.

Ptychography is a scanning version of CDI where one illuminates the sample with a slightly defocused beam (the probe) at multiple locations, collecting a diffraction pattern from each; the illumination spots on the sample overlap with each other, providing the necessary information redundancy to recover the phase from the intensities of diffraction patterns \cite{Rodenburg2004-nw}. In theory, the scan grid of ptychography can be arbitrarily large, which breaks through the field of view limit of full-field CDI. This characteristic has enabled the nanometer-resolution imaging of extensive samples such as microchips (5 \micron{} sample with 4 nm resolution) \cite{Aidukas2024-pk} and brain tissues (10 \micron{} sample with 38 nm resolution) \cite{Bosch2023-fk}.

The forward model of ptychography is formulated as
\begin{equation}
    I_i(\fvec) = \left|\mathcal{F}\left[o(\rvec+\rvec_i)p(\rvec)\right]\right|^2
    \label{eq:ptycho_forward}
\end{equation}
where $\fvec$ and $\rvec$ are reciprocal-space and real-space coordinates, $\mathcal{F}$ is Fourier transform, $o$ and $p$ denote the complex-valued object and probe, and $\rvec$ denotes the probe position of scan point $i$ in the object frame. Phase retrieval algorithms solve for $o$, and sometimes along with $p$ and $\rvec_i$, from the real-valued intensity measurements. The first ptychographic phase retrieval algorithm, the ptychographic iterative engine (PIE), was proposed in \cite{Rodenburg2004-nw}, which is effectively a regularized gradient descent algorithm that finds $o$ as
\begin{equation}
    o = \argmin_{o'}\norm{o'(\rvec+\rvec_i)p(\rvec) - \psi(\rvec)}^2 + \norm{o'(\rvec) - o(\rvec)}_{u(\rvec)}^2
    \label{eq:generic_pie_object_update}
\end{equation}
where $\psi$ is wavefield backpropagated to the object's exit plane after the far-field wavefield's magnitude is replaced with the measured magnitude, $\norm{x}^2_{w} = \sum{w|x|^2}$ is the weighted norm, and $u(\rvec)$ is a regularizer weighting function defined in \cite{Maiden2017-um}. The later proposed extended ptychographic iterative engine (ePIE) \cite{Maiden2009-md} allows the probe to be updated along with the object and adopts a minibatched update scheme that speeds up the convergence. The regularized ptychographic iterative engine (rPIE) algorithm \cite{Maiden2017-um} is a further improvement of PIE and ePIE, which replaces $u(\rvec)$ in Eq.~\ref{eq:generic_pie_object_update} with a more natural form. Beyond the PIE family, various other phase retrieval algorithms exist, which include alternating projection-based difference map (DM) \cite{Elser2003-ov} and relaxed averaged alternating reflections (RAAR) \cite{Luke2005-ps}, second-order methods \cite{Kandel2021-ef, Yang2011-lm}, and methods with look-ahead adaptive step size represented by the least-square maximum likelihood algorithm (LSQML) \cite{Odstrcil2018-ns}. Notably, the LSQML method has been used to deliver reconstructions with remarkable resolution in recent works on the 3D imaging of integrated circuits \cite{Odstrcil2018-eu, Aidukas2024-pk}. 

Despite the rapid development of deep learning (DL)-based phase retrieval approaches, which predict the phase and/or magnitude of the object from one or multiple diffraction patterns without iterative processes \cite{Cherukara2020-yg, Hoidn2023-ed} or with only a few iterations \cite{Gan2024-vi}, analytical phase retrieval algorithms are still the mainstream in the diffraction imaging community. Analytical methods readily transfer across samples and instruments, while DL methods may require continuous fine-tuning or retraining for out-of-distribution samples and probes. They are also more versatile in that they can model and optimize many parameters other than object and probe, such as probe positions \cite{Odstrcil2018-ns} and beam fluctuations \cite{Odstrcil2016-wy}), accounting for the imperfection in these variables and delivering better results. Moreover, analytical methods are more easily extendable to complicated imaging scenarios. For example, while the object is traditionally assumed to be a 2D function, this assumption no longer holds when the object is so thick that the wave diffraction in it is no longer negligible; analytical methods can flexibly model a thick object as a stack of 2D slices with the probe Fresnel propagated between them (multislice modeling) \cite{Maiden2012-qh, Tsai2016-mi}, while DL approaches would have to fine-tune or retrain the model to adapt to the thick-sample scenario, possibly with network architecture changes required. Most importantly, analytical methods are interpretable. When they converge, one can tell if the predicted diffraction intensity $I_i(\fvec)$ agrees with the measurement based on the reconstruction loss, from which one can conclude if the reconstructed images are trustworthy in terms of fidelity. On the other hand, DL methods can hallucinate the presence or absence of features, as there are few mechanisms during inference time to enforce the physics consistency.

This is not to say that analytical methods are immune to unreal artifacts. Under poor or complex conditions, they can also produce reconstructions with artifacts that are potentially worse than DL methods. We illustrate these artifacts with 2 typical examples:
\begin{itemize}
    \item When probe overlap is low and the initial probe guess is bad, grid artifacts appearing like a lattice of dots can emerge in the results. A spatially periodic scan grid worsens these artifacts because of the weakening of solution uniqueness, a phenomenon known as raster grid pathology \cite{Thibault2009-bs}.
    \item In multislice ptychography introduced above, features on a slice could ``seep'' into the reconstructions of adjacent slices at different positions along the beam axis \cite{Maiden2012-qh, Ozturk2018-tv}, particularly when the distances between slices are low. This issue known as crosstalk artifact is a result of the ill-posedness of single-viewing-angle multislice ptychography which does not have a unique solution. Strong prior knowledge is required to disambiguate the slices in order to solve this major problem limiting the axial resolution of multislice ptychography. 
\end{itemize}

Currently, the most effective solution to mitigate raster grid pathology is to use a non-periodic scan grid (such as a Fermat spiral \cite{Huang2014-bp}) and adequate probe overlap, but this is sometimes unattainable especially during joint ptycho-fluorescence microscopy \cite{Deng2015-sq}, where ptychographic diffraction patterns and x-ray fluorescence (XRF) signals are collected at the same time. XRF is a spatial scanning microscopy with lower resolution. As such, when XRF is the main goal of the experiment, the scan points are usually spaced largely to save time and arranged in a rectangular grid due to motor constraints. A computational way to address the grid artifact issue is what we will refer to as ``spectral filtering'', which is based on the observation that the periodic artifacts result in periodic harmonic peaks in the Fourier spectrum of the object; by computing the frequencies of these signals based on the real-space periods of the artifacts and masking them out, one can achieve effective artifact suppression \cite{Huang2017-rz}. This method has gained numerous success in practice, but it is sometimes observed that when the periodicity of the artifacts weakens due to, for example, probe position correction, its effectiveness reduces. For the multislice crosstalk problem, \cite{Huang2019-ou} introduced a method where one generates the low-resolution initial guess without crosstalk features for each slice using the XRF maps of two chemically distinct object slices. The limitation of this approach is that it requires XRF data, and that the slices must not contain the same chemical elements; otherwise, they would not be distinguishable by XRF. A post-processing method was later proposed in \cite{Du2021-kq} which uses deep image prior to remove crosstalk artifacts from phase retrieved images, but it requires a strong spatial gradient correlation penalty to suppress the artifacts effectively, which results in degraded resolution and compromised fidelity. Finally, we note that the community has been long treating grid artifacts, multislice crosstalk, and other types of artifacts separately. A specific solution needs to be devised based on the characteristics of a particular artifact type, and can hardly be transferred to other types. Therefore, the cost of development always increases when new issues are uncovered.

To achieve effective and versatile artifact removal without additional information (\emph{e.g.}, XRF) while maintaining the fidelity and physics consistency of the reconstructions, we consider the alternating direction method of multipliers (ADMM) \cite{Boyd2010-ia}. ADMM is a meta-algorithm that splits a complex problem into multiple subproblems. In every outer iteration of ADMM, each subproblem is solved exactly or approximately with its own solver in a semi-independent fashion, while a dual variable that is updated every epoch encourages the solutions of all subproblems to eventually reach a consensus \cite{Buzzard2017-rx}. This algorithm design brings two desirable properties: first, the mutual consensus pushed by the dual variable effectively maintains the data fidelity of the final solution when there is a subproblem involving model-based reconstruction. Second, the ``bring your own solver'' compatibility allows many existing algorithms to be applied to a specific subproblem. For ptychography, this means one can use well-established methods like ePIE, rPIE, and LSQML for the phase retrieval part. Moreover, ADMM even allows a subproblem to be solved with a no-closed-form, non-gradient-based solver such as a neural network. This strategy, known as plug-and-play (PnP) \cite{Venkatakrishnan2013-ak}, has been widely employed in enhancing the quality of analytical reconstruction methods. 

In a PnP-augmented reconstruction of computational imaging data, the problem is split into a data fidelity subproblem and an image enhancement subproblem. In each outer iteration, the data fidelity subproblem is advanced with several iterations of updates that minimize the distance function between the intensities predicted by the physics image formation model ($I_i$) and those measured in the experiment ($\hat{I}_i$); the image enhancement subproblem is pushed forward by modifying the phase retrieval subproblem's solution with the dual variable and passing it through the image processor; at last, the dual variable is updated. This strategy has demonstrated success in image noise suppression when the image processor is a denoiser \cite{Venkatakrishnan2013-ak}, and in resolution enhancement when the image processor is a deblurring diffusion model \cite{Renaud2024-pa}. Due to the strong presence of the data fidelity subproblem, the final solution still maintains good consistency to measured data. 

However, the artifact removal problem in ptychography is more complicated. First, many neural network models trained to serve as the image processor are small and task specific: they might be trained to remove only one particular type of artifacts that have common and well-defined characteristics, and as such, an image processor trained to remove grid artifacts is often weak to other artifact types. Second, some artifacts, represented by multislice crosstalk, are hard to identify without provisionally provided human knowledge at inference time. For example, given two reconstructed slices that both contain spatially aligned biological cells and nanoparticles, the model may find it hard to tell whether the cells should be exclusive to the first slice and the nanoparticles should be exclusive to the second, or the opposite, or they are just supposed to be mixed. Although crosstalk artifacts are often characterized by their blurriness, the amount of blur varies with many factors, increasing the difficulty of training a widely applicable model to capture and remove blurry features; additionally, these artifacts can also demonstrate characteristics other than blurriness \cite{Du2021-kq}. 

Since scientists often possess extensive prior knowledge of the sample, we can imagine a more versatile image processor that takes natural language inputs from users and specifically removes the features described in the inputs would make PnP algorithms much more flexible and transferable. The development of conditional generative models has made this goal realistic: contrastively trained image/text encoders represented by CLIP \cite{Radford2021-wr} and SigLIP \cite{Zhai2023-et} allow the latent embeddings of texts to be semantically aligned with those of images. When combined with diffusion models through classifier-free guidance \cite{Ho2022-qj}, this allows high-resolution images to be synthesized following text prompts. The mathematical formulation of text-guided diffusion generation is described in prior work \cite{Ho2020-rq, Rombach2021-ke}. Here, we briefly and qualitatively revisit the text-guided generation process of denoising diffusion probabilistic models. A random vector $\xvec_T$ is drawn from a standard Gaussian distribution and is assumed to be at timestep $T$. The vector is then iteratively denoised. At each denoising iteration (timestep), a neural network $\epsilon$ predicts the noise (which is, from a statistics perspective, the score function of the data distribution) in the vector at that timestep both with and without the conditioning of the embeddings of text prompts ($\cvec$), whose outputs are respectively denoted as $\epsilon(\xvec, \cvec, t)$ and $\epsilon(\xvec, t)$. Both predicted noise vectors are combined as $\epsilon(\xvec, t) + s[\epsilon(\xvec, \cvec, t) - \epsilon(\xvec, t)]$, where $s$ is the classifier-free guidance scale. The combined noise is used to denoise $\xvec_t$ following a specific noise scheduler. The denoising operations are repeated until $t = 0$ is reached; for a latent diffusion model, the latent vector at this point, $\xvec_0$, is decoded to pixel-space by a decoder.

On the basis of that, one can also realize text-guided diffusion image editing, where one modifies an existing image instead of generating one from scratch. From a high-level perspective, diffusion image editing differs from direct image generation in that the initial noisy latent vector $\xvec_T$ is not drawn from a Gaussian, but obtained by adding noise to the input image (referred to as the inversion process), which preserves the original image's information. During the denoising process, classifier-free guidance semantically steers the denoising steps, resulting in an output image that structurally resembles the input image but with certain text-described features or concepts altered. Several text-guided image editing approaches employing this or a similar strategy have been proposed, such as DiffEdit \cite{Couairon2022-rd} and Pix2pix-zero \cite{Parmar2023-yb}. To prevent excessive alteration of the original image, some of these methods generate a mask covering only the features described by the text prompts and limit editing to the masked areas. LEDITS++ \cite{Brack2023-cu} is a more recent diffusion image editing method offering several innovations: first, it uses a more efficient noise scheduler requiring fewer inversion and denoising steps; second, it allows multiple concepts to be added or removed; moreover, it computes the editing mask using the intersection between a coarser mask derived from the cross-attention map and a finer mask coming from the editing direction $\epsilon(\xvec, \cvec, t) - \epsilon(\xvec, t)$. These features led to better performance than prior methods \cite{Couairon2022-rd, Parmar2023-yb} and make it a good candidate for the PnP reconstruction framework. Moreover, LEDITS++ has been demonstrated with large foundational diffusion models including Stable Diffusion 1.5 \cite{Rombach2021-ke} and Stable Diffusion XL \cite{Podell2023-sz}, \textit{allowing zero-shot image editing without fine-tuning}. This significantly facilitates the adoption of this technique for high-throughput imaging as a part of a simple and transferable workflow. 

Employing LEDITS++ for the artifact removal subproblem in PnP, we propose an effective, flexible, and fidelity-preserving method for artifact-free ptychographic phase retrieval, featuring:
\begin{itemize}
    \item A PnP framework that pushes for the consensus of data fidelity and artifact removal, ensuring that the reconstructed object is consistent with measured data through the physics model. 

    \item Text-guided artifact removal. Users can instruct the algorithm to remove, for example, ``dot grid'', ``transparent rods'', ``fibers in background'', \emph{etc.}, with natural language, realizing versatile artifact removal without case-specific designs.

    \item The employment of a foundation model (Stable Diffusion 1.5 \cite{Rombach2021-ke}) as the score predictor in reverse diffusion. Though not trained intensively on microscopy images, we show that it is still effective in removing the artifacts in such images.

    \item Drop-in replaceable phase retrieval solver. We show that the data fidelity subproblem in PnP can also be solved using out-of-the-box ptychography reconstruction packages that has been proven to yield fast convergence and good quality, and that has additional features such as partial coherent modeling, probe correction, and probe position correction.
\end{itemize}

We demonstrate our method for grid artifact removal in undersampled rectangular-grid ptychography and for multislice disambiguation, using simulated and experimentally collected measurements. We use LEDITS++ for the artifact removal subproblem, and tested rPIE and LSQML for the data fidelity subproblem. By examining the fidelity of the results by comparing either the peak signal-to-noise ratio (PSNR) with the ground truths or the error between the predicted diffraction patterns and the measurements, we show this method well preserves the consistency with physics and data and delivers high-quality and high-fidelity images. 

\section{Results}

\subsection{Grid artifact removal in ptychographic tomography}

We first test our method with a simulated 3D object that was also used in \cite{Du2020-pv}. Details and images of the object can be found in the supplemental document (Fig.~S1). We begin with the 2D ptychographic reconstruction using the data simulated at the viewing angle of 0$\degree$ of the object. To generate the data, we multiply the complex transmission functions of the object on the $xy$-planes at all 256 integer pixel coordinates along the depth ($z$) direction to get a $256\times256$ 2D projected object, then simulate ptychographic diffraction patterns according to Eq.~\ref{eq:ptycho_forward}. The probe is an experimentally collected disk-shaped spot formed by a zone plate. The diameter of the spot is approximately 50 pixels and the probe's support size is $128\times128$. The scan positions form a rectangular grid with a spacing of 18 pixels in both $x$ and $y$-directions, giving a probe overlapping ratio of $1 - 18 / 50 = 64\%$, which is considered sub-optimal \cite{Bunk2008-to}. Further, to simulate the effect of partial coherence, we use 10 mutually incoherent probe modes \cite{Chen2009-gm} in the simulation. Finally, to create a scenario where the probe function is not exactly known at reconstruction, we magnify the true probe by 5\% to generate the initial guess for the probe. 

\begin{figure}
    \centering
    \includegraphics[width=0.95\linewidth]{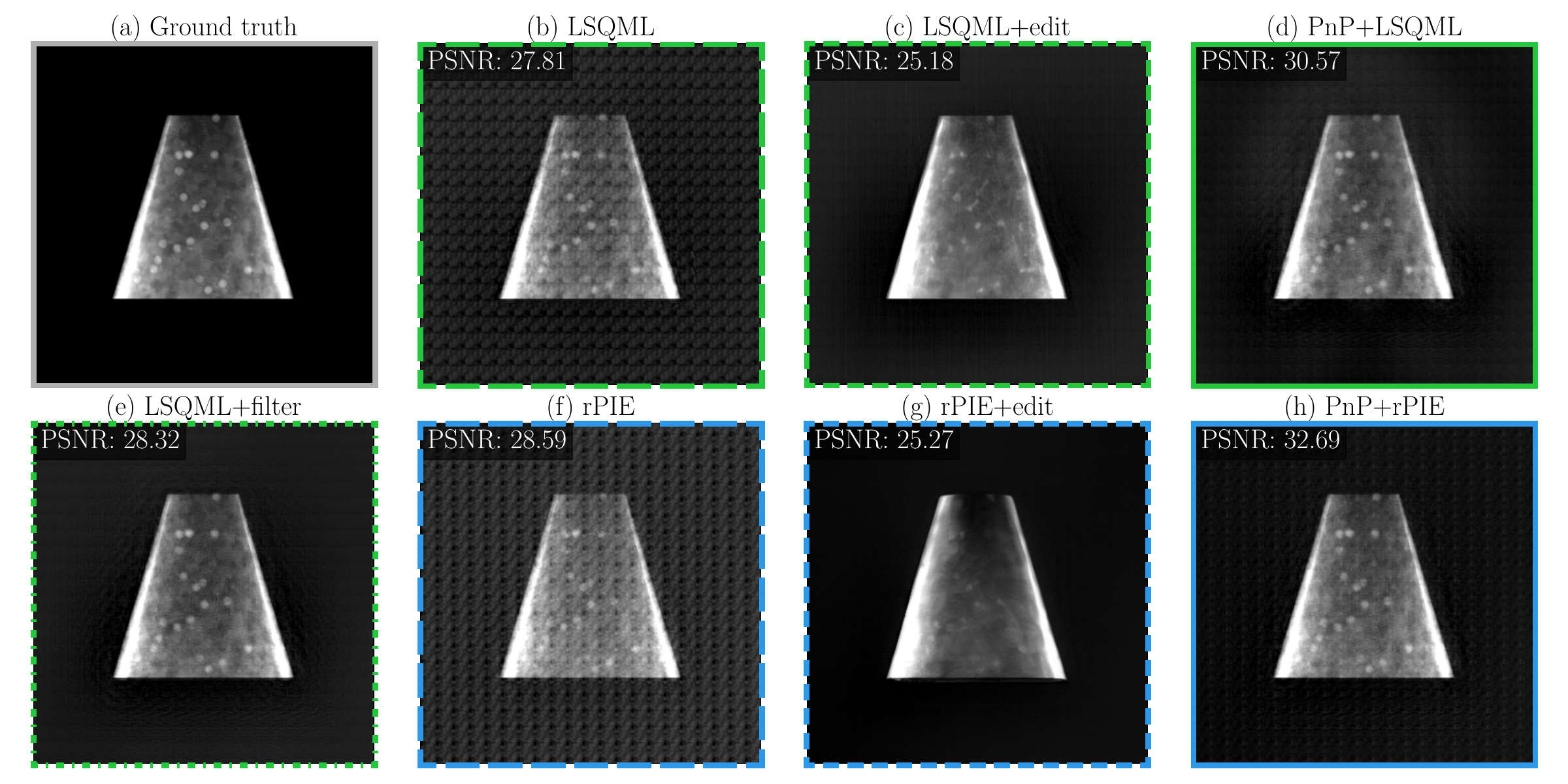}
    \caption{True and reconstructed phase images of the tapered tube object using projection data simulated at the viewing angle of 0$\degree$. All images are the phase of the object. LEDITS++'s editing prompt is set to remove ``dot grid''. The subplots show (a) the ground truth, and the reconstructions with (b) vanilla LSQML, (c) LSQML followed by image editing, (d) our method that combines reconstruction and image editing in the PnP framework, and (e) LSQML with spectral filtering. (f-h) are the counterparts with rPIE as the ptychographic solver. The PSNR of the reconstruction compared to the ground truth are indicated at the top left corner of each subplot. The colors and line styles of the image borders are visual aids that indicate the ptychographic solver and the enhancement method.}
    \label{fig:cone_angle0_recons}
\end{figure}

We reconstructed this simulated data using LSQML and rPIE for 1000 iterations also with 10 probe modes and with probe correction enabled. Still, the results are not ideal due to low overlap and grid pathology. As Fig.~\ref{fig:cone_angle0_recons}(b) and (f) show, both reconstructions (phase images) are degraded by periodic grid artifacts. 

Before assessing our method, we first evaluate the effect of using LEDITS++ image editing as a ``post-processing'' technique, where it edits the image at the end of vanilla ptychographic phase retrieval. The results with LSQML and rPIE, shown in Fig.~\ref{fig:cone_angle0_recons}(d) and (g), show that while the artifacts appear to be effectively removed, the PSNRs drop even lower than those of the vanilla solvers. A closer examination with the ground truth reveals that the structures of the object are distorted and smeared, a result of structurally poorly constrained diffusion generation. Without the constraint of data and physics, fidelity is sacrificed in this edit-after-reconstruct approach. 

We then reconstructed the data with our method. The number of inversion and inference steps for image editing was set to 100. The editing prompt of LEDITS++ was set to remove ``dot grid'', and the classifier-free guidance scale for editing was set to 5. The ADMM proximal penalty $\admmpenalty$ was set to $10^{-5}$ and the relaxation was set to $0.8$. Image editing (Eq.~\ref{eq:image_editing}) was run after every 250 iterations of the ptychography solver (Eq.~\ref{eq:ptycho_solve} and \ref{eq:ptycho_proximal}), repeating for 4 outer iterations, giving the same total of 1000 ptychographic reconstruction iterations. Fig.~\ref{fig:cone_angle0_recons}(e) and (h) show the reconstructed images are clean, with the grid artifacts significantly suppressed.

To make an additional baseline case for comparison, we also examine the results of spectral filtering. This operation was implemented by calculating the locations of the harmonic peaks in the Fourier spectrum of the object corresponding to the periodicity of the grid artifact, and setting the $5\times 5$ neighborhood around each peak to zero. This processing was run every 100 epochs throughout the reconstruction (not including the last epoch). As indicated in Fig.~\ref{fig:cone_angle0_recons}(e), spectral filtering (with LSQML) is also effective in suppressing the artifacts, and since the filtering is done in an alternating manner with phase retrieval, structural fidelity is preserved as well, yet the PSNR (28.32) is lower than our method (30.57). While the filtering method works reasonably well on this simulated dataset, we will show a corner case of it on an experimental dataset in the next section. 

\begin{figure}
    \centering
    \includegraphics[width=0.95\linewidth]{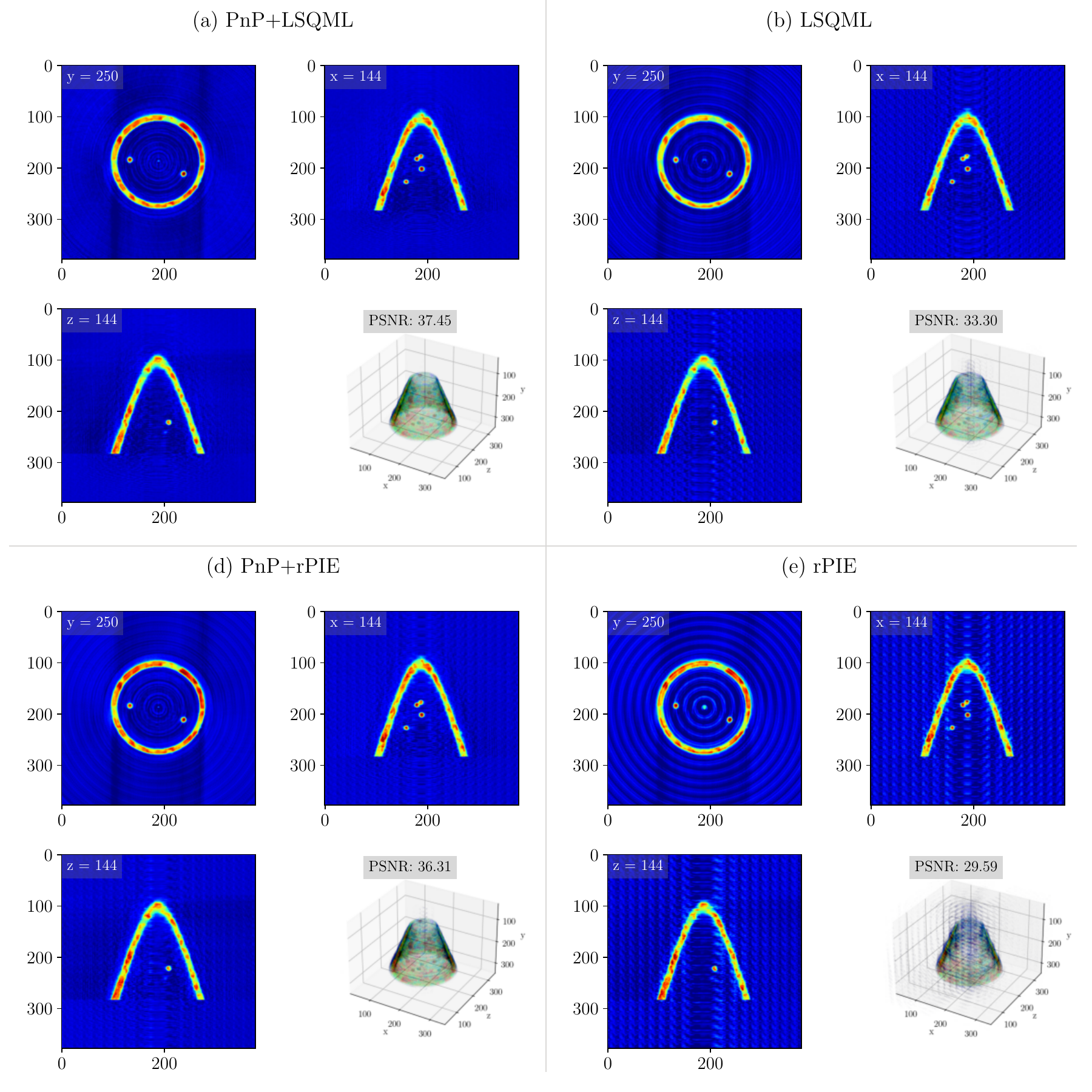}
    \caption{The tomography reconstructions of the object. Results with projection images obtained using (a) our method with LSQML, (b) vanilla LSQML, and (c, d) the counterparts with rPIE are shown. For each method, we show the three orthogonal slices at the same positions as the ground truth in Fig.~S1. The PSNR values on the top of each 3D rendering are calculated between the 3D reconstruction obtained with projection images retrieved using the tested method and that obtained with the projections of the ground truth 3D object.}
    \label{fig:cone_tomo_recons}
\end{figure}

To further verify the artifact suppression and structural fidelity of our method, we simulated the 2D ptychography data at 180 viewing angles in the range between 0$\degree$ and 180$\degree$, and reconstructed the data at each angle with the same parameters as the 0$\degree$-reconstruction shown above. We then performed a tomographic reconstruction with the phase-retrieved projection images at all viewing angles using the algebraic reconstruction technique (ART) \cite{Gordon1970-art} implemented in TomoPy \cite{Gursoy2014-oz} to obtain the object in 3D. Examining the accuracy of the internal structures of the 3D reconstruction provides a more solid assessment on the fidelity of the 2D phase retrieval at each angle. The images shown in Fig.~\ref{fig:cone_tomo_recons} show the 3D objects reconstructed with the 2D projection images obtained using various approaches, including our method with LSQML as the ptychographic solver (a), vanilla LSQML (b), and the counterparts with rPIE (c, d). For both ptychographic solvers, the PnP method clearly reduces the artifacts that come from grid pathology in 2D projections. These artifacts exhibit as concentric rings in the horizontal slices ($y = 250$) and unequally spaced short lines in the vertical slices ($x = 144$ and $z = 144$). This improvement is also apparent in the 3D renderings, where the results of vanilla methods are degraded by an array of rings in the 3D space, while those of our methods are much cleaner. We also show the 3D PSNR between the reconstruction objects and the ground truth. Since our focus is the 2D phase retrieval, we exclude the error due to tomography reconstruction by using the ART reconstruction of the projections of the ground truth object, rather than the 3D ground truth itself, for PSNR calculation. The PSNR values also indicate superior performance of our method compared to vanilla algorithms as well as LSQML with spectral filtering (36.96, images not shown).

\begin{figure}
    \centering
    \includegraphics[width=0.95\linewidth]{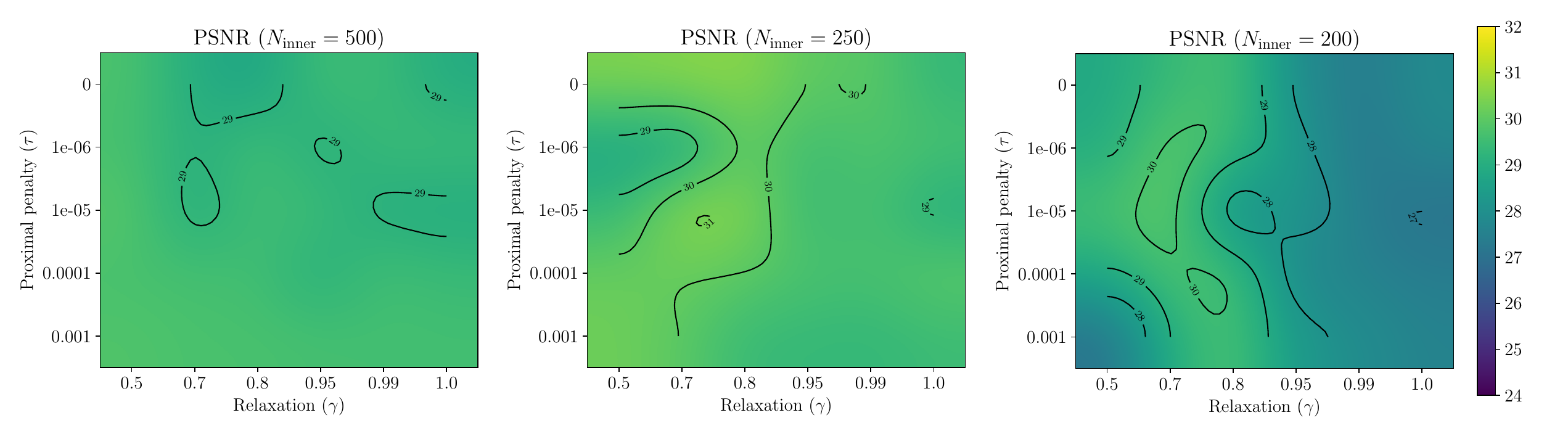}
    \caption{PSNR of the 2D reconstructed phase at 0$\degree$ viewing angle of the tube object with different algorithm parameters. The three plots are respective for a number of phase retrieval inner iterations of 500, 250, and 200. The total number of phase retrieval iterations is always 1000, so the number of outer iterations of PnP are 2, 4, and 5. In Each plot, PSNR is plotted against the proximal penalty ($\admmpenalty$) and the relaxation factor ($\gamma$). PSNR values outside the tested data points are interpolated.}
    \label{fig:cone_angle0_metric_vs_params}
\end{figure}

For further insights on the effects of critical algorithm parameters such as the penalty ($\admmpenalty$), relaxation factor ($\gamma$), and the number of phase retrieval iterations before each image editing run ($\npr$), we conducted a series of experiments using different combinations of these parameters. We set the total number of phase retrieval iterations to 1000, and varied $\npr$ among 500, 250, and 200, giving 2, 4, and 5 outer iterations respectively. $\admmpenalty$ was varied among 0 and the range between $10^{-6}$ and $10^{-3}$, and $\gamma$ was varied between the range of 0.5 to 1.0. The results are shown in Fig.~\ref{fig:cone_angle0_metric_vs_params}, where the PSNR values out of the tested data points are interpolated. $\npr = 250$ results in the best overall PSNR: an $\npr$ that is too high reduces the number of outer iterations and the occurrences of image editing, limiting the effectiveness of artifact suppression, while an   $\npr$ too low causes the data fidelity subproblem to be solved too inexactly, undermining the accuracy of the solution. The relaxation factor $\gamma$ has a sweet spot of around 0.7 to 0.8. The relaxation allows the solution of the data fidelity subproblem to be mixed into the outcome of image editing, enhancing the accuracy, but an excessively low $\gamma$ causes the artifact-containing data fidelity solution to dominate, undoing the effect of image editing. The proximal penalty $\admmpenalty$ determines how strongly the data fidelity subproblem's solution $o$ should be kept close to the artifact removal subproblem's solution $v$. When $\admmpenalty$ is set to 0, the ptychographic solver operates independently when solving the data fidelity problem. We should note that this still gives reasonably good performance as although the data fidelity subproblem becomes unconstrained, it is still pushed closer to the artifact-free local minimum by the image editor because the edited image is used in the initial guess for the ptychographic solver in the next outer iteration. However, a non-zero penalty enhances the constraint coming from the artifact removal subproblem. A too-high penalty, on the other hand, makes the artifact removal subproblem's solution dominate, which hurts data fidelity. 

\subsection{Grid artifact removal with experimental ptychography data}

This section demonstrates our method on ptychography data experimentally collected at the Advanced Photon Source. The probe is also a zone-plate formed disk-shaped spot with a diameter of about 50 pixels. A rectangular scan grid with a spacing of approximately 18.8 pixels in $x$ and $y$ is used, giving a relatively low probe overlapping ratio of 62\%. 

Like the simulated case, we test both rPIE and LSQML as the ptychographic solver. We used 10 probe modes to account for partial coherence, and enabled the gradient-based probe position correction feature in the reconstruction tool (Pty-Chi) \cite{Du2025-nq} to reduce errors in recorded probe positions. These configurations were applied both when the solvers were used as is and when they were used in the PnP method. 

\begin{figure}
    \centering
    \includegraphics[width=0.9\linewidth]{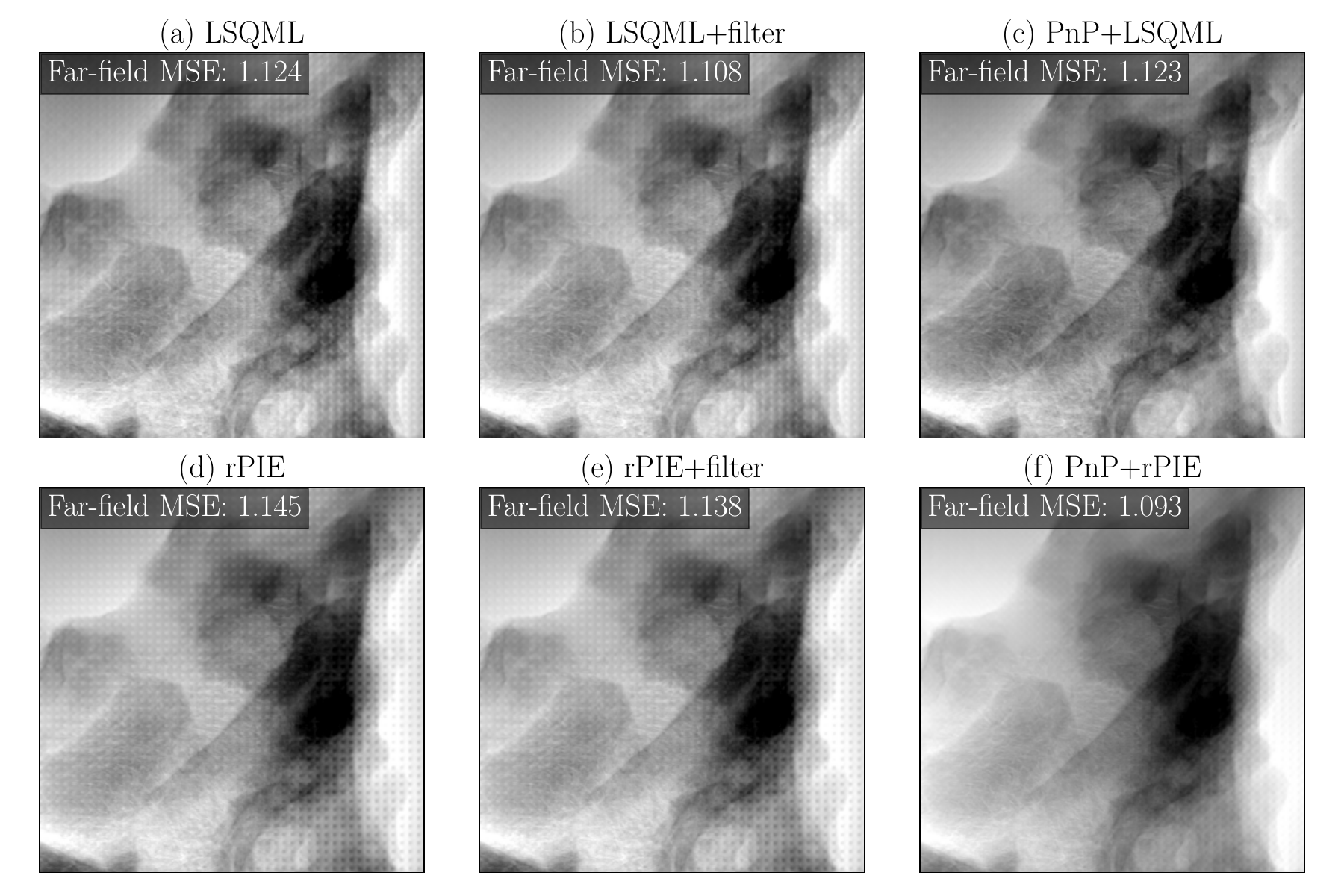}
    \caption{The reconstructed phase of the experimental ptychography data using various methods. (a) is obtained using vanilla LSQML. (b) is obtained using LSQML with spectral filtering. (c) is obtained using our method with LSQML as the ptychographic solver. (d-f) are the counterparts with rPIE as the ptychographic solver. The mean squared errors of the magnitude of the far-field diffraction patterns are shown on the top left corner, indicating that our method delivers results with reduced grid artifacts while maintaining a data consistency on par or even better than vanilla approaches.}
    \label{fig:bnp_recons}
\end{figure}

For vanilla algorithms, the low overlap and rectangular grid resulted in increased difficulty to reconstruct the data. A total of 1000 phase retrieval iterations were used for both ptychographic solvers and both vanilla and PnP cases. The reconstructed phase images with both vanilla LSQML and rPIE are degraded by grid artifacts as shown in Fig.~\ref{fig:bnp_recons}(a) and (d). 

Even with spectral filtering, the grid artifact did not improve significantly, as shown in Fig.~\ref{fig:bnp_recons}(b) and (e). We have also observed during the reconstruction that although spectral filtering indeed reduced the artifacts right after it was applied (particularly at early iterations), the artifacts quickly re-emerged. This is partly because probe position correction was enabled during reconstruction, which reduces the periodicity of the artifacts and broadens the harmonic peaks corresponding to the artifacts in the Fourier spectrum of the object, making the filtering approach less effective. The inaccurate initial probe further complicated the reconstruction.

The reconstructions with the PnP approach were performed with $\admmpenalty = 10^{-4}$, $\gamma = 1$, $\npr = 100$, 100 inference steps, and with LEDITS++ set to remove ``dot grid'' with a classifier-free guidance scale of 7. To further enhance data fidelity, we limited image editing to only run in the first 7 outer iterations, so the last 300 phase retrieval iterations were run consecutively. The results shown in Fig.~\ref{fig:bnp_recons}(c) and (f) have the grid artifacts reduced particularly in the center and top-left regions of the images. Since this is an experimental dataset without ground truth, we are unable to compute the PSNRs of the reconstructions, but we can still examine the consistency of the results with the measured data by calculating the mean squared error between the magnitudes of the far-field diffraction patterns predicted by the physics model and the measurements. For LSQML, the error of PnP is on par with that of the vanilla algorithm, while for rPIE, the error of PnP is even lower than vanilla phase retrieval. We explain this lower error by considering the cause of grid pathology described in \cite{Thibault2009-bs}: assuming $o(\rvec)$ and $p(\rvec)$ are ideal object and probe solutions of a ptychography problem without artifacts, then $f(\rvec)o(\rvec)$ and $f^{-1}(\rvec)p(\rvec)$ are also solutions if and only if $f(\rvec) = f(\rvec + \rvec_i)$ for all $i$'s, because this is the only way to satisfy
\begin{equation}
    f(\rvec + \rvec_i)o(\rvec + \rvec_i) \cdot f^{-1}(\rvec)p(\rvec) = o(\rvec + \rvec_i)p(\rvec) \qquad \forall i \in 1, \cdots, N_\mathrm{DP}.
\end{equation}
For any $f(\rvec)$, $f(\rvec)o(\rvec)$ is a local solution of the problem, and its degeneracy is increased by low probe overlap and periodic scan grid. Applying image editing is equivalent to transforming $f(\rvec)$ close to a constant function, eliminating most of the non-unique solutions and pushing the solution to the true global minimum.

\subsection{Crosstalk artifact removal in multislice ptychography}

Crosstalk is a result of ill-posedness of single-view multislice ptychography, where one seeks an object function composed of multiple 2D slices stacked along the beam axis. In the forward model of multislice ptychography, wave modulated by an upstream slice is Fresnel propagated to the next slice. The exit wave $\psi_i$ is given by the following written in linear algebraic form:
\begin{equation}
    \psi_i = \mathrm{diag}(\boldsymbol{o}_{\rvec+\rvec_i}^{N_s})\prod_{j=1}^{N_s-1}\left[ \mathbf{F}_{d_j}\mathrm{diag}(\boldsymbol{o}_{\rvec+\rvec_i}^j) \right]\boldsymbol{p}
\end{equation}
where $\mathbf{F}_{d_j}$ is the Fresnel propagator over distance $d_j$, and $N_s$ is the number of slices. The ability to  distinguish different slices relies on the incommutability of $\mathrm{diag}(\boldsymbol{o}_{\rvec+\rvec_i}^j)$ due to the introduction of the Fresnel propagation operator. When the inter-slice spacing $d_j$ is small, the Fresnel propagator approaches the identity operator. That is,
\begin{equation}
    \lim_{d_j \rightarrow 0~ \forall~ 1\le j \le N_s}\psi_i = \prod_{j=1}^{N_s}\left[\mathrm{diag}(\boldsymbol{o}_{\rvec+\rvec_i}^{j}) \right]\boldsymbol{p}.
\end{equation}
Since $\mathrm{diag}(\boldsymbol{o}_{\rvec+\rvec_i}^j)$ are now element-wise multiplied, a unique solution for each slice no longer exists. Through the removal of crosstalk artifacts, our goal is to push the solution of each slice towards the solution we desire and away from local minima.

\begin{figure}
    \centering
    \includegraphics[width=0.95\linewidth]{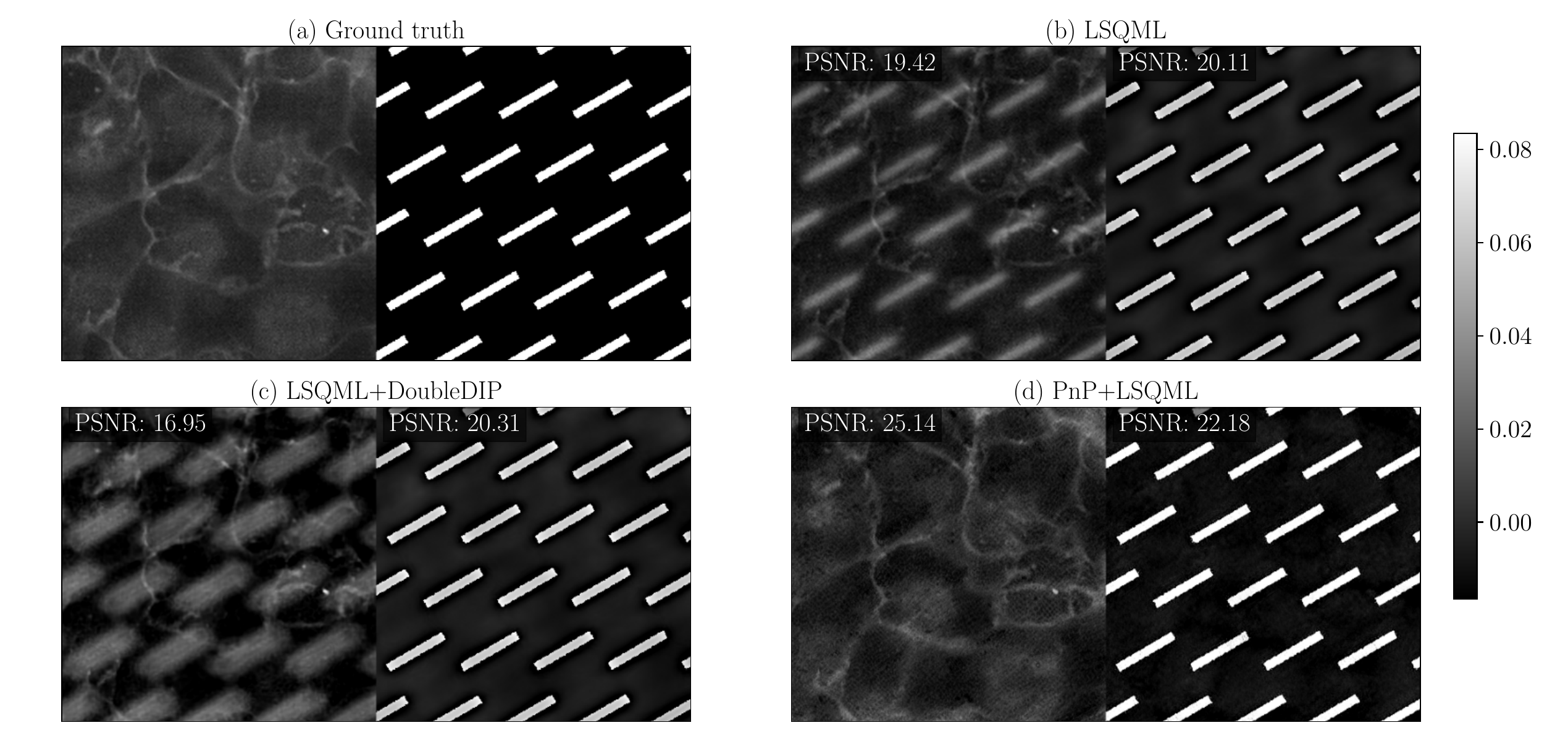}
    \caption{Ground truth and reconstructions of the simulated multislice object. (a) The ground truth phase images of both slices. (b) The reconstructed phase images with vanilla LSQML. (c) The reconstructions of vanilla LSQML post-processed by DoubleDIP. (d) The reconstructions obtained using PnP with the LSQML solver. The per-slice PSNR values are shown at the top-left corners. Colorbar on the right applies to all images.}
    \label{fig:multislice_recons}
\end{figure}

For this experiment, we created a simulated two-slice object with overlapping yet distinct features: the first slice is a fluorescence image of cells obtained through \emph{scikit-image} \cite{scikit-image}, and the second slice contains an array of rods mimicking nano-imprinted structures. The ground truth phase images are shown in Fig.~\ref{fig:multislice_recons}(a). To generate the diffraction data, we assumed a beam energy of 10 keV and a lateral pixel size of 10 nm. This gives a depth of field (DOF) of 4.35 \micron{} \cite{Jacobsen2019-iq}. We then assume the distance of both slices to be 10 \micron{} or 2.3 times the DOF.

We use LSQML as the ptychographic solver. The reconstructions of vanilla LSQML, shown in Fig.~\ref{fig:multislice_recons}(b), exhibit obvious crosstalk especially on the first slice where the rods that should have been exclusively on the second slice appear in the first slice. On the second slice, the contrast of the rods is worse, and the background is less uniform. 

Our next baseline result is obtained using the post-processing method introduced in \cite{Du2021-kq} based on double deep image prior (DoubleDIP) \cite{Gandelsman2018-od}. Given the images with crosstalk artifacts, this method employs 3 neural networks that map random noise to supposed clean images of both slices as well as two mixing ratios. The generated images are mixed to match the crosstalk-containing images. The tendency of neural network to generate natural images with low information entropy serves as a prior constraint (deep image prior \cite{Lempitsky2018-ee}) that drives the neural networks to generate crosstalk-free images that reproduce the crosstalk-corrupted images when mixed. However, the performance of this method for our test case is poor: as shown in Fig.~\ref{fig:multislice_recons}(c), the crosstalk on the first slice is merely smeared instead of suppressed. The smaller spacing-to-DOF ratio is a factor contributing to the failure of the DoubleDIP method in this case: in \cite{Du2021-kq}, the slice spacing of the first test case is 3.6 times the DOF, larger than the dataset used in this paper, while the second test case is added by additional XRF data. Smaller slice spacing results in stronger coupling of features across the slices, making the crosstalk more difficult to remove.

For the PnP approach, we use 200 inference steps, a classifier-free guidance scale of 6, $\admmpenalty = 0.01$, $\gamma = 0.95$, $\npr = 100$, and $N_\mathrm{outer} = 10$ giving a total of 1000 phase retrieval iterations. The removal prompt for LEDITS++ is ``transparent rods'' for the first slice, and ``fibers in background'' in the second slice. Additionally, to prevent the drift in image brightness and contrast, we run an image statistics matching after each image editing: we first calculate the absolute difference between the images before and after editing, and threshold it to obtain a mask giving pixels that are not significantly modified by editing. We then offset and scale the edited image so that it has the same mean and standard deviation as the pre-editing image within the masked region. For this experiment, the matching was only done to the first slice, and the mask threshold is set to 0.5. The reconstructed phase images are shown in Fig.~\ref{fig:multislice_recons}(d), where no obvious crosstalk is seen. Comparing the first slice with the ground truth, we conclude that the structures of the object are recovered with good fidelity. 

\section{Discussion}

The image editing model, along with the phase retrieval solver, are the two main pillars of our algorithm. Good characteristics of the image editing model are close prompt following and minimal alterations to the input image other than the concepts described by the prompts. LEDITS++ is good at preserving non-artifact features given that the classifier-free guidance scale and editing threshold are set properly as a result of the fine-grained masks. With the built-in text encoder of Stable Diffusion 1.5, the image editor also follows editing prompts well as demonstrated in the above cases where it was asked to remove ``dot grid'', ``transparent rods'' and ``fibers in background'', but we occasionally observed inconsistent performance from run to run. With the advancement of large, extensively trained text encoders and text-image alignment techniques, more promising candidates as the text-steered image editor are emerging. For example, the GPT Image 1 model of OpenAI, which is integrated in the ChatGPT models \cite{OpenAI2024-pq}, has demonstrated superior prompt following and good preservation of non-editing features without tuning extra parameters. While GPT Image 1 is closed-weight, we expect open-weight models to catch up in this domain, at which point they will greatly benefit the large-scale deployment of generative model-based methods like ours. 

In this work, we use an off-the-shelf, pretrained foundation model as the underlying score predictor. A foundation model is powerful enough to generalize to various types of artifacts and samples without any training whatsoever, either pre-training or fine-tuning to a domain task. Hence, through the use of a foundation model, the production applicability of our method is greatly enhanced, and the development cycle is significantly accelerated, especially when constrained by resources and data. However, we do recognize the limitation of using foundation models, especially those trained mostly on natural, artistic images and general, common text corpora instead of those encountered more often in microscopy imaging. Training or fine-tuning a generative model on the distribution of microscopy images, and a text encoder that well aligns terms and concepts in microscopy with image features takes painstakingly curated data and heavy computation, but would be highly beneficial to image enhancement in the context of scientific imaging and is a promising direction that we will direct our future efforts to. 

The introduction of image editing in the PnP method adds to the computation time of reconstruction. Since the time for phase retrieval and image editing scales with different factors, it is hard to give a general estimate of the time added by image editing with regard to phase retrieval. We list the time consumed by phase retrieval for $\npr$ iterations and that consumed by one image editing run, along with the relevant data sizes and settings, in Table \ref{tab:walltime}. Both walltimes listed roughly add up to the time taken by one outer iteration of PnP as the time for dual update is relatively negligible. In our test cases, the time consumed by image editing ranges from 38\% to 82\% of phase retrieval. While this is a substantial addition, we note that when larger diffraction patterns are used in phase retrieval for higher spatial resolution, the time for phase retrieval scales up roughly in $O(N\log N)$ because of the increased data size for fast Fourier transform, while the time taken by image editing is unaffected assuming the object size stays the same. Also, when phase retrieval needs more iterations to converge, $\npr$ increases, making the time taken by image editing less significant in a relative sense. The method, however, is more memory-bound. With a $512\times 512$ single-slice object, the Stable Diffusion 1.5 model we used took about 30 GB more GPU memory with the score predictor of the model running in 16-bit precision and the variation autoencoder running in 32-bit precision for numerical stability. 
Smaller generative models may help relieve the memory requirement at the cost of generalizability. When the algorithm is used by multiple end users at a large facility (\emph{e.g.}, beamlines in a synchrotron light source), it is possible to host the model on a powerful, centralized server, with the individual users sending image editing tasks to that server and fetching results from it. One may also resort to cloud providers for either GPUs or publicly hosted generative models which atomizes the cost of computing resources on an as-needed basis.

\begin{table}
    \small
    \centering
    \begin{tabular}{cccccccccc}
    \hline
    \textbf{Test case} & 
    \makecell{\textbf{Diffraction}\\ \textbf{patterns}} & 
    \makecell{\textbf{Probe}\\ \textbf{size}} & 
    \makecell{\textbf{Probe}\\ \textbf{modes}} & 
    \makecell{\textbf{Object}\\ \textbf{size}} & 
    \makecell{\textbf{Object}\\ \textbf{slices}} & 
    $\mathbf{\npr}$ & 
    \makecell{\textbf{Time for }\\ $\mathbf{\npr}$\\ \textbf{PR iters}\\ \textbf{(LSQML, s)}} & 
    \makecell{\textbf{Inference}\\ \textbf{steps}} & 
    \makecell{\textbf{Time for}\\ \textbf{image editing}\\ \textbf{(one run, s)}}  \\
    \hline
    Tube & 484 & 128 & 10 & 512 & 1 & 250 & 16 & 100 & 6  \\
    Experimental & 2209 & 128 & 10 & 1024 & 1 & 100 & 32 & 100 & 20 \\
    Multislice & 1600 & 128 & 10 & 512 & 2 & 100 & 34 & 200 & 28 \\
    \hline
    \end{tabular}
    \caption{Data sizes, relevant settings, and computation walltimes of the 3 case studies.}
    \label{tab:walltime}
\end{table}

\section{Methods}

\subsection{Plug-and-play}

We start by introducing the fundamentals of ADMM in the scenario of the problem of ptychographic reconstruction with artifact removal. The problem is formulated as
\begin{equation}
    \min_{o,\Theta,v} f(o, \Theta) + g(v) \qquad \text{s.t.}~o = v
    \label{eq:constrained_objective}
\end{equation}
where $f$ is the objective function of the phase retrieval data fidelity subproblem which quantifies the mismatch between predicted intensities $I_i(\fvec)$ and measured intensities $\hat{I_i}(\fvec)$, $g$ is the objective function of the artifact removal subproblem, $o$ and $v$ are the unknown variables representing the reconstructed object in both subproblems respectively, and $\Theta$ is the set of other unknowns including the probe $p$, probe positions $\{\rvec_i|i = 1, \cdots, N\}$, \emph{etc.} Eq.~\ref{eq:constrained_objective} can be converted to an unconstrained optimization problem using Lagrangian multipliers, giving the following augmented Lagrangian \cite{Venkatakrishnan2013-ak}:
\begin{equation}
    \mathcal{L} = f(o, \Theta) + g(v) + \frac{\admmpenalty}{2}\norm{o - v + u}^2 - \frac{\admmpenalty}{2}\norm{u}^2
    \label{eq:lagrangian}
\end{equation}
where $u$ is the Lagrangian multiplier or the scaled version of the dual variable. Eq.~\ref{eq:lagrangian} is solved by alternatingly minimizing $\mathcal{L}$ with regards to $o$, $v$, and $u$. Further, we adapt the relaxation mechanism \cite{Boyd2010-ia} to modify the solution of the artifact removal subproblem, $v$, by mixing it with the solution of the data fidelity subproblem, $o$, to further promote the consensus of both subproblems. This gives the following steps in an outer iteration of ADMM \cite{Venkatakrishnan2013-ak}:
\begin{eqnarray}
        o &\leftarrow& \argmin_o \left\{ f(o, \Theta) + \frac{\admmpenalty}{2}\norm{o - v + u}^2 \right\} \label{eq:admm_o}\\
        v &\leftarrow& \argmin_v \left\{ g(v) + \frac{\admmpenalty}{2}\norm{o - v + u}^2 \right\} \label{eq:admm_v} \\
        v &\leftarrow& \gamma v + (1 - \gamma)o \\
        u &\leftarrow& u + o - v \label{eq:admm_u}
\end{eqnarray}
where $\gamma$ is the relaxation factor. The plug-and-play approach \cite{Venkatakrishnan2013-ak} proposes the replacement of the inner minimization for a certain subproblem with a solver $D$ that non-iteratively maps the input to the solution, \emph{i.e.}, Eq.~\ref{eq:admm_v} is replaced with
\begin{equation}
    v \leftarrow D(o + u).
    \label{eq:image_editing}
\end{equation}
We further propose that the data fidelity subproblem of Eq.~\ref{eq:admm_o} can also be solved using an off-the-shelf ptychographic solver. While these solvers are typically designed only to minimize $f(o, \Theta)$ without the proximal term $\frac{\admmpenalty}{2}\norm{o - v + u}^2$, we recognize that many ptychography solvers are iterative, and after each inner iteration of the ptychographic solver, we perform a gradient descent step to enforce the proximal term. With this, Eq.~\ref{eq:admm_o} is implemented as an iterative process, where in each iteration we run
\begin{eqnarray}
    o, \Theta &\leftarrow& F(o, \Theta, \hat{I}) \label{eq:ptycho_solve} \\
    o &\leftarrow& o-\admmpenalty\bigl(o-v+u\bigr) \label{eq:ptycho_proximal}
\end{eqnarray}
where $F$ is the operator that executes one iteration of the ptychographic solver. Combining the designs above, we summarize our algorithm in Algorithm \ref{alg:pseodocode}.

Since the ptychographic data used in our experiments are mostly stronger in phase contrast and weaker in absorption contrast, the reconstructed phase images are usually cleaner and the magnitude images are usually noisier and have bad contrast. Therefore, for our test cases, image editing is only applied on the phase, and the complex object after editing is formed by combining the edited phase and a constant magnitude. The magnitude is left for the ptychographic solver to find out. 

We use LEDITS++ \cite{Brack2023-cu} as the single-shot artifact removal problem solver, $D$. Stable Diffusion 1.5 \cite{Rombach2021-ke} is used as the underlying diffusion model. 

\begin{algorithm}[t]
\caption{Plug-and-Play ptychography with diffusion-based artifact removal}
\label{alg:pseodocode}

\KwIn{%
    Initial complex object image $o_0$; initial guesses of other unknowns $\Theta_0$; measured diffraction patterns $\hat{I}$; ADMM dual variable $u\leftarrow 0$; Auxiliary variable $v\leftarrow 0$; ADMM penalty $\admmpenalty>0$; relaxation factor $\gamma\in[0,1]$; Ptychography iterator $F(\cdot)$; diffusion image editor $D(\cdot)$; Outer/inner iteration counts $N_{\mathrm{outer}}$, $\npr$; text description of the artifacts to remove $c$
}
\KwOut{Reconstructed image $x$ free of text-described artifacts}

$o\leftarrow o_0$\;
$\Theta\leftarrow \Theta_0$\;
\tcp{Outer ADMM loop}
\For{$\text{epoch}=0,\dots,N_{\mathrm{outer}}-1$}{
  \lIf{$\text{epoch}>0$}{$o\leftarrow v-u$}%
  \tcp{Ptychography reconstruction step}
  \For{$k=0,\dots,\npr-1$}{
        $o, \Theta\leftarrow F(o, \Theta, \hat{I})$\;
        \If{$\text{epoch}>0$}{
            $o\leftarrow o-\admmpenalty\bigl(o-v+u\bigr)$\;
        }
  }
  \tcp{Image editing step}
  $v'\leftarrow D(o+u)$\;
  $v\leftarrow \gamma\,v + (1-\gamma)\,o$\;
  \tcp{Dual variable update}
  $u\leftarrow u + o - v$\;
}
\Return{$o$}\;
\end{algorithm}

\subsection{Implementation}

The algorithm was implemented in Python. For LEDITS++, we used its implementation in Hugging Face Diffusers \cite{VonPlaten2022-diffusers}. For the ptychographic solver, we use the ptychography reconstruction package Pty-Chi \cite{Du2025-nq} developed at the Advanced Photon Source, Argonne National Laboratory, which offers a collection of reconstruction algorithms including rPIE \cite{Maiden2017-um} and least-square maximum likelihood (LSQML) \cite{Odstrcil2018-ns}. All computations were performed on a Supermicro 740GP-TNRT server using an NVIDIA H100 GPU. 

\subsection{Sample preparation and data collection}

The sample used in the beamline experiment was based on a magnesium potassium phosphate-based (MKP) cement developed at Argonne National Laboratory. The procedure for synthesizing the phosphate cement has been discussed in detail elsewhere \cite{Singh2001-pumpable, Singh1998-phosphate}. 

The imaging experiment was conducted at the Bionanoprobe beamline at the Advanced Photon Source \cite{Chen2014-jj}. The sample was affixed to the apex of a sharp tungsten tip with a two-component adhesive. This arrangement allowed the specimen to be analyzed over the full 360$\degree$ range without interference from a sample holder. The sample was then loaded into the vacuum chamber. A monochromatic X-ray beam at 10 keV was focused using a stacked zone plate of 70 nm outermost zone width and 160 \micron{} diameter. The sample was positioned about 1 mm downstream of the focal point, resulting in a beam size of $\sim$1.3 \micron{} on the particle. Ptychographic scan was performed using an on-the-fly scanning scheme over a 45 \micron{} $\times$ 30 \micron{} field of view, with a step size of 0.5 \micron{} in both horizontal and vertical directions. Far-field diffraction patterns were then collected using an in-vacuum Eiger2 X 1M detector placed 2.06 m downstream of the sample, with an exposure time of 20 ms for each diffraction pattern.


\section{Conclusion}

We developed a method to suppress and remove artifacts in ptychographic phase retrieval results. We use the plug-and-play (PnP) strategy to combine model-based phase retrieval and generative image editing, pushing the solution towards a consensus of both problems and maintaining data fidelity. We employed a text-guided image editing method using a foundational diffusion model, which allows users to specify the artifacts to remove in natural language and results in the broad applicability to various artifact types. We also allow the phase retrieval problem in the PnP framework to be solved by most existing ptychographic reconstruction tools, enhancing its flexibility and making the additional features of these tools accessible. We demonstrated our method on the removal of grid artifacts due to periodic scan points and undersampling, as well as crosstalk artifacts in multislice ptychography, using both simulated and experimental data, and showed superior performance compared to vanilla reconstruction algorithms without any fine-tuning or re-training. This work not only advances the field of ptychographic phase retrieval but also demonstrates the potential of integrating foundational generative models into scientific imaging workflows. Through future works on training a more microscopy science-oriented foundation model, we will continue to improve the effectiveness and robustness of our method.

\section*{Acknowledgements}

This research used resources of the Advanced Photon Source, a U.S. Department of Energy (DOE) Office of Science user facility at Argonne National Laboratory, and is based on research supported by the U.S.~DOE Office of Science-Basic Energy Sciences, under Contract No.~DE-AC02-06CH11357.

\section*{Code availability}

The code developed for the described method is available at \url{https://github.com/AdvancedPhotonSource/aether}.

\printbibliography

\end{document}


\maketitle

\section{Simulated 3D object}

\begin{figure}
    \centering
    \includegraphics[width=0.6\linewidth]{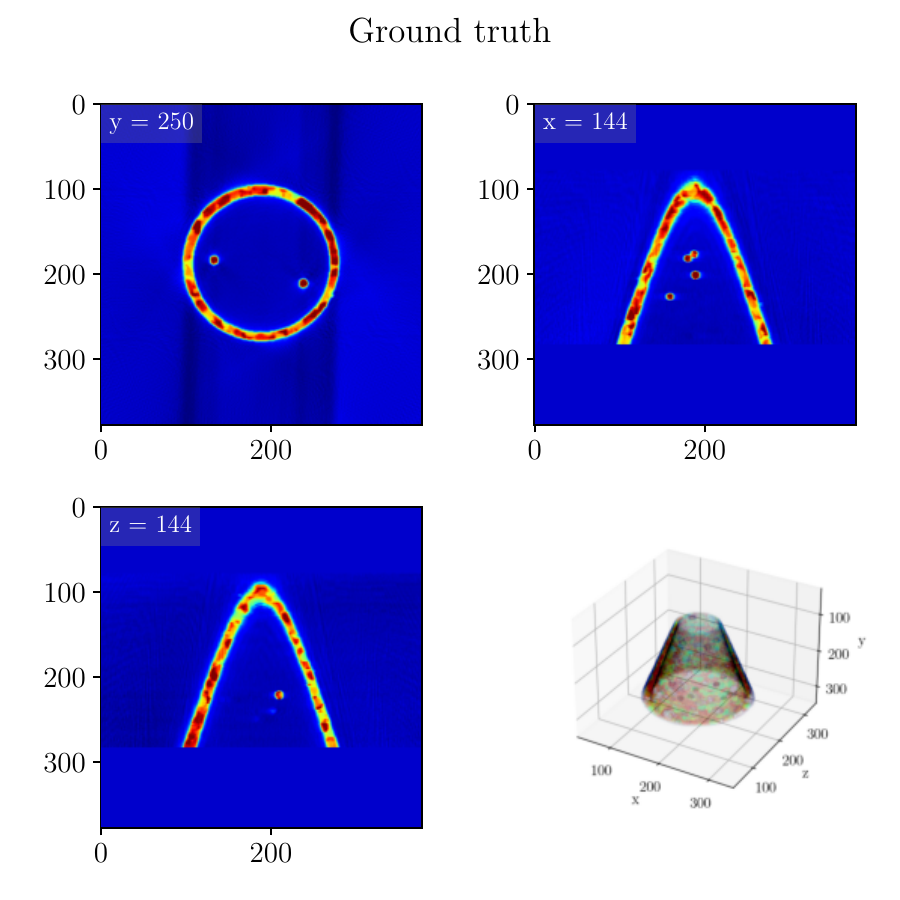}
    \caption{The simulated 3D object. To facilitate later comparison with the tomography reconstructions obtained with various methods, the images shown here are also tomographically reconstructed from the projection images of the ground truth. The three 2D images show 3 slices of the object, which include the $xz$-cross section at $y = 250$, the $yz$-cross section at $x = 144$, and the $xy$-cross section at $z = 144$. The two vertical slices are slightly off-centered to reveal a larger area of the wall of the tube and more particles inside of it. A transparent 3D rendering is also shown in the figure. }
    \label{fig:cone_gt}
\end{figure}

The simulated 3D object used in the first case study is described here. The object is a hollow tube that tapers towards the top. The wall of the tube is randomly patterned with circles with varying absorptions and phase shifts. Small spherical particles are randomly scattered on the outer wall, and several larger particles are contained inside the tube. The object contains $256\times256\times256$ voxels. Three orthogonal cross sections and a 3D rendering of the object are shown in Fig.~\ref{fig:cone_gt}. To allow direct comparison with the tomography reconstructions from the phase-retrieved projection images obtained with various methods, images shown here are also from the 3D tomography reconstruction with projection images generated by summing the phase of the ground truth object.

